\documentclass[letter]{aa} 
\usepackage{graphicx}
\usepackage{txfonts}
\usepackage{natbib}
\bibpunct{(}{)}{;}{a}{}{,}
\usepackage{amssymb}
\usepackage{xcolor}
\usepackage{lipsum}
\usepackage{textcomp}
\usepackage[colorlinks=true,linkcolor=blue,citecolor=blue,
bookmarks=true,bookmarksopen=true,bookmarksnumbered=true,draft=false]{hyperref}
\usepackage{enumerate}

\newcommand{\snr}{SNR 1987A}
\newcommand{\xmm}{{\it XMM-Newton}}
\newcommand{\chandra}{{\it Chandra}}

\begin{document}

   \title{XMM-Newton observations of SNR 1987A. II.\,\thanks{Based on
observations obtained with \xmm, an ESA science mission with instruments and
contributions directly funded by ESA Member States and NASA.}}

   \subtitle{The still increasing X-ray light curve and the properties of Fe K
lines}

	\titlerunning{XMM-Newton observations of SNR 1987A. II. X-ray light curve
and Fe K lines}

   \author{P. Maggi  \inst{1}
	\and F.   Haberl \inst{1}
	\and R.   Sturm  \inst{1}
	\and D.   Dewey  \inst{2}
	}

   \institute{Max-Planck-Institut f\"ur extraterrestrische Physik, Postfach
	1312, Giessenbachstr., 85741 Garching, Germany\\ \email{pmaggi@mpe.mpg.de}
	\and
	MIT Kavli Institute, Cambridge, MA 02139, USA
    }

   \date{Received 19 October 2012 / Accepted 6 November 2012}

  \abstract
  {}
   {We report on the recent observations of the supernova remnant \object{SNR
1987A} in the Large Magellanic Cloud with \xmm. Carefully monitoring the
evolution of the X-ray light curve allows to probe the complex circumstellar
medium structure observed around the supernova progenitor star.}
   {We analyse all \xmm\ observations of \snr\ from January 2007 to December
2011, using data from the EPIC-pn camera. Spectra from all epochs are extracted
and analysed in a homogeneous way. Using a multi-shock model to fit the
spectra across the 0.2--10 keV band we measure soft and hard X-ray fluxes with 
high accuracy. In the hard X-ray band we examine the presence and properties of
Fe  K lines. Our findings are interpreted in the framework of a
hydrodynamics-based model.}
  {The soft X-ray flux of \snr\ continuously increased in the recent years.
Although the light curve shows a mild flattening, there is no sudden break as 
reported in an earlier work, a picture echoed by a revision of the \chandra\ 
light curve. We therefore conclude that material in the equatorial ring and
out-of-plane \ion{H}{ii} regions are still being swept-up. We estimate the
thickness of the equatorial ring to be at least $4.5 \times 10 ^{16}$ cm
(0.0146 pc). This lower limit will increase as long as the soft X-ray flux has
not reached a turn-over. We detect a broad Fe K line in all spectra from 2007 to
2011. The widths and centroid energies of the lines indicate the presence of a
collection of iron ionisation stages. Thermal emission from the hydrodynamic
model does not reproduce the low-energy part of the line (6.4--6.5 keV),
suggesting that fluorescence from neutral and/or low ionisation Fe might be
present.
}
   {}


   \maketitle

\defcitealias{2011ApJ...733L..35P}{P11}
\defcitealias{2012ApJ...752..103D}{D12}
\defcitealias{helder2012_tmp}{H12}
\section{Introduction}
\label{introduction}
\snr\ is one of the most studied objects in the southern sky. Because of its
location in the Large Magellanic Cloud (LMC) at a distance of 50 kpc, it can be
resolved at radio, optical, and even X-ray wavelengths. X-ray observatories such
as {\it ROSAT}, \chandra\ and \xmm\ have frequently observed \snr, offering a
unique opportunity to follow the early evolution of an SNR.

The most important feature of the soft X-ray light curve (between 0.5 and 2 keV)
has been the upturn observed about 6\,000 days after the explosion
\citep{2005ApJ...634L..73P}, interpreted as the beginning of the interaction of
the blast wave with an ``equatorial ring'' (ER) of denser material around the
progenitor star \citep[see for instance Fig.\,7 in][]{2002ApJ...572..209S}.
The ER is likely to have been formed by the interaction between
the stellar winds emitted by the progenitor star during its red supergiant and
blue supergiant phases \citep[e.g.][]{1995ApJ...452L..45C}, although
binary merger models also exist to explain such a structure
\citep[e.g.][]{2007Sci...315.1103M}. Monitoring the evolution of the X-ray light
curve allows to probe the structure of the ring and to constrain the late stages
of the progenitor. \citet[][hereafter \citetalias{2012ApJ...752..103D}]{
2012ApJ...752..103D} presented simple hydrodynamic models that reproduce the
soft and hard X-ray light curves; the models show the soft X-ray flux behaviour
for both the case where the forward shock has left the ER and the case where the
ER is still being shocked (the ``thin'' and ``thick'' cases in their Figure 12).

\citet[][hereafter \citetalias{2011ApJ...733L..35P}]{ 2011ApJ...733L..35P}
presented recent \chandra\ observations (up to September 2010). Owing to the
apparent flattening of the soft X-ray light curve, these authors concluded that
\snr\ had reached a new evolutionary phase, where the blast-wave has passed the
main body of the ER and is now interacting with matter with a decreasing density
gradient. However, \citet[][hereafter
\citetalias{helder2012_tmp}]{helder2012_tmp} used the revised ACIS calibration
to analyse all \chandra\ observations (including new ones, up to March 2012).
They concluded that the sudden break reported by
\citetalias{2011ApJ...733L..35P} was only a calibration effect.

In this Letter, we present our latest \xmm\ monitoring observations
(Sect.\,\ref{observations}). We focus first on the evolution of the X-ray flux
(Sect.\,\ref{xray_light_curve}), then report the detection of Fe K lines
(Sect.\,\ref{Fe_K_lines}). Discussion and conclusions are given in
Sect.\,\ref{Discussion}.

\begin{table*}[t]
\caption{Details of the XMM-Newton EPIC-pn observations}
\label{table_observations}
\centering
\begin{tabular}{c c c c c c c c c c}
\hline\hline
\noalign{\smallskip}
ObsId & Obs. start date & Age \tablefootmark{a} &  Filter  & Total\,/\,filtered
exp. time \tablefootmark{b} & Flux (0.5--2 keV) & Flux (3--10 keV) &
$\dot{F_\mathrm{X}}$ \tablefootmark{c}\\
 & & (days) & & (ks) & \multicolumn{2}{c}{
$\left(10^{-13}\right.$ erg\,s$^{-1}$\,cm$\left.^{-2}\right)$} & (\%) \\
\noalign{\smallskip}
\hline
\noalign{\smallskip}
0406840301 & 2007 Jan 17 & 7267 & Medium & 106.9\,/\,81.9 & 
33.94$^{+0.46}_{-0.49}$ & 4.25$^{+0.72}_{-0.79}$ & --- \\
\noalign{\smallskip}
0506220101 & 2008 Jan 11 & 7626 & Medium & 109.4\,/\,91.0 & 
43.76$^{+0.47}_{-0.62}$ & 5.45$^{+0.70}_{-1.45}$ & 31.1 \\
\noalign{\smallskip}
0556350101 & 2009 Jan 30 & 8013 & Medium & 100.0\,/\,84.0 & 
53.05$^{+0.61}_{-0.57}$  & 6.54$^{+0.69}_{-1.62}$ & 14.3 \\
\noalign{\smallskip}
0601200101 & 2009 Dec 12 & 8328 & Medium & 89.9\,/\,89.8  & 
60.09$^{+0.61}_{-0.52}$ & 7.82$^{+0.57}_{-1.34}$ & 31.2 \\
\noalign{\smallskip}
0650420101 & 2010 Dec 12 & 8693 & Medium & 64.0\,/\,61.7  & 
66.91$^{+0.87}_{-0.90}$ & 9.32$^{+0.74}_{-1.57}$  & 11.3 \\
\noalign{\smallskip}
0671080101 & 2011 Dec 02 & 9048 & Medium & 80.6\,/\,70.5  & 
71.88$^{+0.48}_{-0.70}$ & 11.31$^{+0.66}_{-1.82}$ & 10.5 \\
\noalign{\smallskip}
\hline
\end{tabular}
\tablefoot{Fluxes are given with 3$\sigma$ errors (99.73\,\% C.L.).
\tablefoottext{a}{Number of days since the explosion of SN\,1987A.}
\tablefoottext{b}{Total and useful (filtered) exposure times, after removal of
high background intervals.}
\tablefoottext{c}{Increase rate in \% since previous measurement, normalised to
one year.}
}
\end{table*}

\section{Observations and data reduction}
\label{observations}
The X-ray photons from \snr\ were amongst the first \xmm\ detected in
January 2000. This observation and two others from 2001 and 2003 were analysed
by \citet{2006A&A...460..811H}. We then started a yearly monitoring of \snr,
from Januray 2007 \citep{2008ApJ...676..361H} to December 2011. The
high-resolution Reflection Grating Spectrometer (RGS) data taken up to
January 2009 are presented in \citet{2010A&A...515A...5S}.

Here we homogeneously (re-)analyse all observations from 2007 to 2011, three of
which have not been published so far. We mainly use data from the EPIC-pn camera
\citep{2001A&A...365L..18S}, operated in Full-Frame mode with medium filter.
Details of the observations are listed in Table\,\ref{table_observations}. We
processed all observations with the SAS\footnote{Science Analysis Software,
\url{http://xmm.esac.esa.int/sas/}} version 11.0.1. We extracted spectra from a
circular region centered on the source, with a radius of 25\arcsec. The
background spectra were extracted from a nearby point source-free region common
to all observations. We selected single-pixel events (\texttt{PATTERN} = 0) from
the pn detector. We rebinned the spectra with a minimum of 20 counts per bin in
order to allow the use of the $\chi ^2$-statistic. Non-rebinned spectra were
used with the C-statistic \citep{1979ApJ...228..939C} for the study of Fe K
lines (see Sect.\,\ref{Fe_K_lines}), because of the limited photon statistics
above 6 keV. The spectral fitting package XSPEC \citep{1996ASPC..101...17A}
version 12.7.0u was used to perform the spectral analysis.

\section{X-ray light curve}
\label{xray_light_curve}
To measure the X-ray flux of \snr, we fitted the EPIC-pn spectra with a
three-component plane-parallel shock model (called vpshock in XSPEC, where the
prefix ``v'' indicates that abundances can vary), using \texttt{neivers} 2.0.
This is the same model as the one used by \citetalias{2012ApJ...752..103D} for
\chandra\ and \xmm\ spectra, with a fixed-temperature component ($kT=1.15$ keV),
although we did not use a Gaussian smoothing. This model gives slightly better
fits than when using a two-component
model\citep[\emph{e.g.}][]{2004ApJ...610..275P,2008ApJ...676..361H}. The high
and low-temperature components are believed to originate from the interaction of
the shocks with uniform material and denser clumps in the ER, respectively
\citepalias{2012ApJ...752..103D}. Another interpretation is that the
high-temperature component comes from plasma shocked a second time by a
reflected shock \citep{2006ApJ...645..293Z}.

For elemental abundances, we followed the same procedure as in
\citet{2006A&A...460..811H}\,: N, O, Ne, Mg, Si, S and Fe abundances were
allowed to vary but were the same for all observations, whereas the He, C, Ar,
Ca and Ni abundances were fixed. The systemic velocity of \snr\ \citep[286
km\,s$^{-1}$, \emph{e.g.}][]{2008A&A...492..481G} was taken into account by
choosing the redshift accordingly.

For the absorption of the source emission, we included two photoelectric
absorption components, one with N$_{H\mathrm{\ Gal}} = 0.6 \times 10^{21}\ 
\mathrm{cm}^{-2}$ (fixed) for the Galactic foreground absorption
\citep{1990ARA&A..28..215D} and another one with N$_{H\mathrm{\ LMC}}$ (free in
the fit) for the LMC. Metal abundances for the second absorption component are
fixed to the average metallicity in the LMC \citep[\emph{i.e.}, half the solar
values,][] {1992ApJ...384..508R}. All spectra share the same N$_{H\mathrm{\
LMC}}$.

We simultaneously fitted the six spectra using energies between 0.2 and 10 keV.
For consistency with the detection of Fe K lines (see Sect.\,\ref{Fe_K_lines}),
we included an additional (Gaussian) line to the model for the spectra obtained
after 2007. The central energies and widths of the lines were fixed to the
values found in the detailed analysis (Sect.\,\ref{Fe_K_lines}). Only the
normalisation of each line was let free.

\begin{figure}[t]
	\centering
	\includegraphics[angle=-90,width=\hsize]
{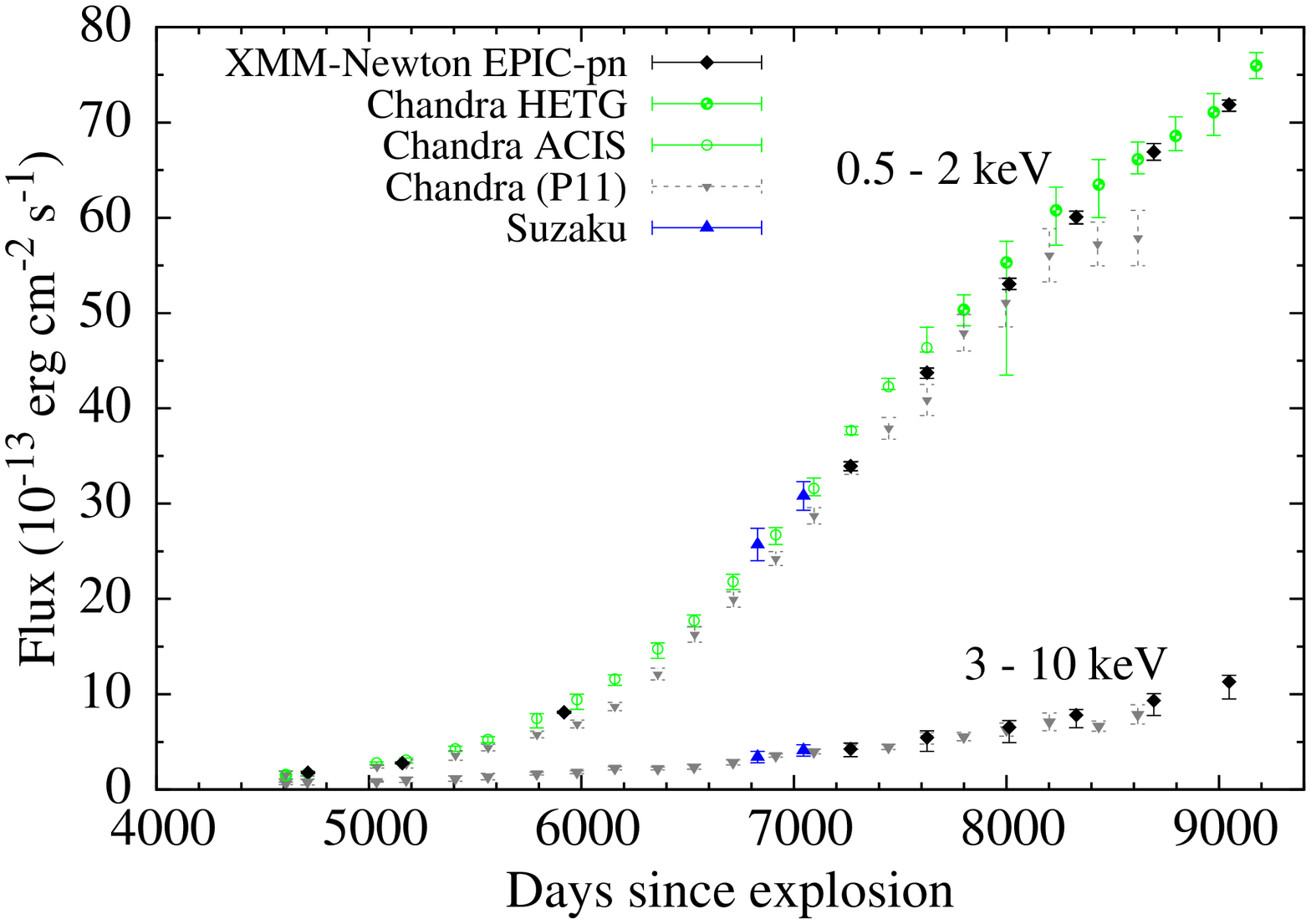}
	\caption{Light curve of \snr\ in the soft and hard X-ray ranges. \xmm\ data
points (black diamonds) after day 7000 are given with 99.73\,\% C.\,L. error
bars. Updated \chandra\ measurements \citepalias[(with 68\,\%
C.\,L. errors,][]{helder2012_tmp} and those based on the older calibration
\citepalias[with 90\,\% C.\,L. errors,][]{2011ApJ...733L..35P} are shown in
green and gray, respectively. Suzaku measurements \citep[blue
triangles]{2009PASJ...61..895S} are also shown.}
\label{fig_light_curve}
\end{figure}

The fit was satisfactory, with $\chi ^2 = 4125.06$ for 3443 degrees of freedom.
Although detailed spectral fits are not the focus of this study, we found that
\emph{(i)} the best-fit N$_{H\mathrm{\ LMC}}$ was $3.09_{-0.08}^{+0.07} \times
10^{21}$ cm$^{-2}$, corresponding to a total absorption column of $3.7
\times 10^{21}$ cm$^{-2}$, somewhat higher than found using the grating
instruments aboard \chandra\ and \xmm\
\citep{2010A&A...515A...5S,2006ApJ...645..293Z},
\emph{(ii)} the abundance pattern is in line with the one reported by
\citet{2010A&A...515A...5S} and \citetalias{2012ApJ...752..103D},
\emph{(iii)} the temperature of the cool component increased only slightly from
0.34 to 0.38 keV, while its normalisation remained constant, and
\emph{(iv)} the temperature of the hot component is always in excess of 3.5
keV, and its normalisation and contribution steadily increased.

\begin{figure*}[!ht]
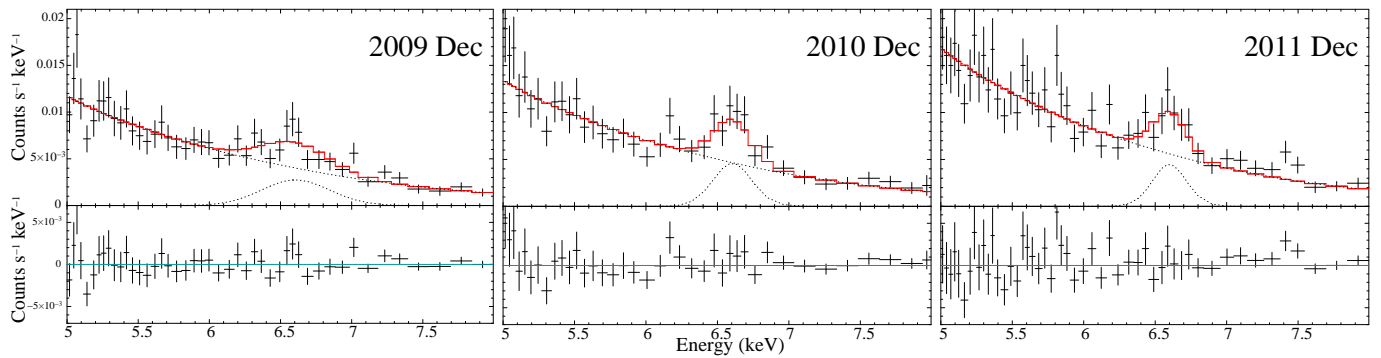

	\centering
\includegraphics
[bb= 80 14 584 700,clip,angle=-90,width=0.34560\hsize]{Felines2009b.ps}
\includegraphics
[bb= 80 90 574 700,clip,angle=-90,width=0.30860\hsize]{Felines2010.ps}
\includegraphics
[bb= 80 90 584 700,clip,angle=-90,width=0.30860\hsize]{Felines2011.ps}
\caption{Recent EPIC-pn spectra of \snr\ in the 5--8 keV range, showing the Fe K
lines. The model used (red) is the sum of two components\,: a bremsstrahlung
continuum and a Gaussian, shown by dotted lines. The bottom panels show the fit
residuals. All panels have the same scale. Note that for plotting purpose only,
adjacent bins are rebinned in order to have a significant ($\geq 5\sigma$)
detection in each rebinned channel. The feature seen at $\sim$7.4 keV in the
2011 spectrum is an instrumental artefact and not a \mbox{Ni K line.}}
\label{fig_Felines}
\end{figure*}

We measured the soft (0.5--2 keV) and hard (3--10 keV) fluxes at all epochs,
using the XSPEC \texttt{flux} command. We list the results, with $3\sigma$
uncertainties (99.73\,\% confidence level, C.\,L.) in
Table\,\ref{table_observations}. Note that as customary for \snr\ we give
\emph{absorbed} fluxes, so comparisons between various observatories are easier
(because they do not depend on the column densities obtained from the fit). The
fluxes up to January 2009 are fully consistent with the results from
\citet{2008ApJ...676..361H} and \citet{2010A&A...515A...5S} which used the same
data.

We included these results in the X-ray light curve shown in
Fig.\,\ref{fig_light_curve}. Older \xmm\ fluxes are taken from
\citet{2006A&A...460..811H}. We show the \chandra\ measurements with the old
calibration \citepalias{2011ApJ...733L..35P} and the newest one
\citepalias{helder2012_tmp}. We also add the results from Suzaku observations
\citep{2009PASJ...61..895S}.

Within their respective errors, \xmm, \chandra, and Suzaku measured soft and
hard X-ray X-ray fluxes which agree very well. \citetalias{2011ApJ...733L..35P},
using the ACIS calibration available at that time, stated that the soft X-ray
flux from \snr\ has been nearly constant after day $\sim 8000$. Obviously \xmm\
observed a source that was \emph{not} constant, although we do observe a mild
flattening of the light curve. The increase rates of the soft X-ray flux (last
column of Table\,\ref{table_observations}) vary from one year to another,
showing that the evolution of the X-ray flux is not smooth. One should therefore
be cautious when claiming a steepening or flattening of the light curve and wait
for a longer baseline.

The discrepancy between \chandra\ and \xmm\ measurements after day 8000 is
reconciled by \citetalias{helder2012_tmp}, using the revised \chandra\
calibration. They concluded that the apparent break in the soft X-ray light
curve \citepalias{2011ApJ...733L..35P} was mainly due to build-up of
contamination on the ACIS optical blocking filters.

\section{Fe K lines}
\label{Fe_K_lines}

The superior high-energy effective area of \xmm\ ($\sim 900$ cm$^{-2}$ at 6.4
keV \emph{vs.} $\sim 200$ cm$^{-2}$ for \chandra) allows the study of Fe K lines
with the EPIC cameras, at energies between 6.4 and 6.7 keV, \emph{i.e.} out of
the range covered by RGS. \citet{2008ApJ...676..361H} noted a possible detection
of an Fe K$\alpha$ line in the spectrum obtained in 2007, but the insufficient
statistics precluded a more detailed analysis. In the coadded spectra from 2007
to January 2009, \citet{2010A&A...515A...5S} identified a line at $6.57 \pm
0.08$ keV.

We analysed the presence and properties of Fe K lines in all the monitoring
observations (Table\,\ref{table_observations}). We fitted the non-rebinned
spectra with a bremsstrahlung continuum and a Gaussian line, making use of the
C-statistic to take into account the limited number of counts in each bin. We
performed \emph{F-tests} to evaluate the significance of the line in each
observation\,: we found a detection more than $3\sigma$ (respectively
$4\sigma$) significant in the data from 2008 and 2011 (respectively December
2009 and 2010). The January 2009 observation, having a slightly shorter exposure
due to longer high background activity periods, still yields a $2\sigma$
detection. We found only a marginal ($1\sigma$) detection in the 2007 spectrum,
in agreement with previous studies.

We show the lines in the three unpublished spectra in Fig.\,\ref{fig_Felines}.
The plasma temperatures of the bremsstrahlung continua range from  $kT = 2.75$
to 3 keV, and the emission measures follow the increasing trend of the hard
X-ray flux. Line properties are given in Table\,\ref{table_Felines} for all
observations except the one from 2007.

\begin{table}[t]
\caption{Fe K line properties.}
\label{table_Felines}
\centering
\begin{tabular}{l c c c c}
\hline
\hline
\noalign{\smallskip}
 Epoch & $E_{\mathrm{line}}$ & $\sigma$-width & Flux & EW \\
 & (keV) & (eV) & ($10^{-6}$ ph\,cm$^{-2}$\,s$^{-1}$) & (eV) \\
\noalign{\smallskip}
\hline
\noalign{\smallskip}
2008 Jan&6.58$^{+0.05}_{-0.07}$& 46 $(< 146)$&1.07$^{+0.57}_{-0.46}$&174 \\
2009 Jan&6.55$^{+0.15}_{-0.14}$&125 $(< 433)$&0.94$^{+1.50}_{-0.65}$&169 \\
2009 Dec&6.63$^{+0.11}_{-0.09}$&229 $(< 424)$&2.64$^{+1.69}_{-1.78}$&432 \\
2010 Dec&6.61$^{+0.06}_{-0.06}$&105 $(< 227)$&2.51$^{+1.39}_{-0.99}$&344 \\
2011 Dec&6.61$^{+0.06}_{-0.06}$& 83 $(< 178)$&2.08$^{+1.01}_{-0.86}$&238 \\
\noalign{\smallskip}
\hline
\end{tabular}
\tablefoot{Central energy, $\sigma$-width, total photon flux and equivalent
width (EW) of the Gaussian used to characterise the Fe K feature in the spectra
of \snr. We give 90\,\% C.\,L. errors.
}

\end{table}

No evolution of the line central energy is found within the uncertainties, but
our spectra suffer from limited statistics. The energy of the Fe K line depends
on the ionisation stage of iron, increasing from 6.4 keV for \ion{Fe}{ii},
to 6.7 keV for \ion{Fe}{xxv} \citep{1986LNP...266..249M,2004ApJS..155..675K}.
The spectral resolution is only $\sim 160$ eV \citep{2001A&A...365L..18S}, so we
are not able to resolve possible contributions from different Fe ions present in
the X-ray emitting plasma. Therefore, the large measured widths of the lines in
our spectra are most likely a sum of lines (which might be Doppler-broadened)
from several Fe ions, convolved with the instrumental response of the camera.
The weighted average of the emission-line centroids (6.60 $\pm 0.01$ keV) and
the typical widths ($\sim 100$ eV) indicate the presence of ionisation stages
from \ion{Fe}{xvii} to \ion{Fe}{xxiv}. This is consistent with the detection of
lines from \ion{Fe}{xvii} to \ion{Fe}{xx} in the RGS spectra
\citep[][]{2010A&A...515A...5S}, and the detection of a
\ion{Fe}{xxii}--\ion{Fe}{xxiii} blend in the \chandra\ High Energy Transmission
Grating spectra \citep{2008ApJ...676L.131D}.

The flux in the iron line indicates an increase around day 8000, and a decrease
afterwards. However, given the large (statistical) errors in the flux
measurements, we cannot exclude the possibility that the flux of the line
remained constant in the last three years. Since the hard continuum steadily
increased during that period of time, the equivalent width of the line decreased
in the last observations (Table\,\ref{table_Felines}).

\begin{figure}[t]
	\centering
	\includegraphics[bb= 81 17 570 700, clip, angle=-90,width=\hsize]
{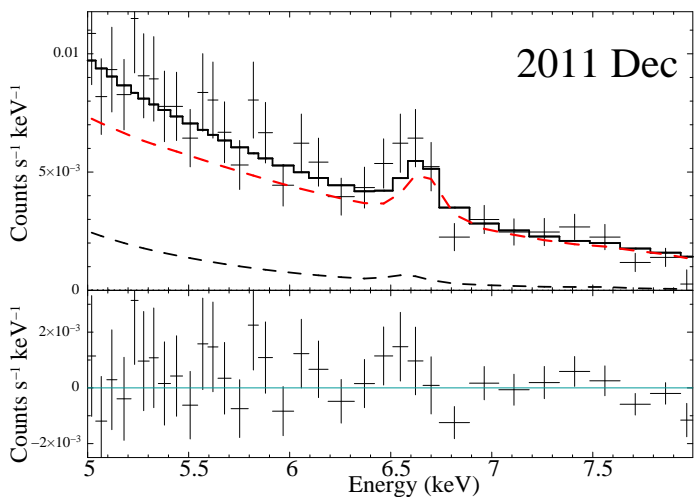}
	\caption{Details of the Fe K lines region of the spectrum from December
2011. The model used is the three-shock components model described in
Sect.\,\ref{xray_light_curve}, including emission from Fe below He-like ions
and without a Gaussian line. The ``hot'' component (red dashed line) dominates
the continuum and line spectrum but does not account for an emission excess
around 6.55 keV.}
\label{fig_3vpshockFelines}
\end{figure}

\section{Discussion and Conclusions}
\label{Discussion}

Our monitoring campaign with \xmm\ since 2007 is ideally suited to follow the
evolution of \snr\,:
\emph{(i)} we used the same intrument setting (observing mode, filter, read-out
time);
\emph{(ii)} we extracted the spectra in the same regions and
\emph{(iii)} we used the same model for all the spectra. 
The high throughput of \xmm\ results in a high statistical quality of our
spectra. This allows high-confidence flux measurements with relatively small
errors and free of cross-calibration issues due to different observing modes.

Our light curve shows a continuous increase of the soft X-ray flux of \snr,
indicating that no turn-over has been reached yet and that the blast wave is
still propagating into dense regions of the ER. To further constrain the
thickness of the ER, given the continuous increase of the soft X-ray flux, we
use the ``$2 \times 1$D'' hydrodynamical model from
\citetalias{2012ApJ...752..103D}. The recent \xmm\ and \chandra\ measurements
point towards a thickness of at least $4.5 \times 10 ^{16}$ cm (0.0146 pc) for
the ER, and each year of continued flux increase requires an additional ER
thickness of $\sim 0.53 \times 10^{16}$ cm (0.0017 pc).

The high-energy collective power of \xmm\ allows us to detect and characterise
the Fe K lines from \snr. We find that the energies and the widths of the lines
imply the presence of a collection of ionisation stages for iron.
To investigate which model component is most responsible for the Fe K lines, we
use the best-fit three-shock model (switching off the Gaussian line) and using
the NEI version 1.1, as it includes low ionisation stages (below He-like
ions), which allows to probe the whole range of energy between 6.4 and 6.7 keV
as function of $kT$ and $\tau$ \citep[see Fig.\,3 in][]{2009ApJ...693L..61F}.
As expected, we find that only the high-temperature component significantly
contributes to the hard continuum and the line. When including emission from
ions below He-like iron, the shapes and fluxes of the lines from the
high-temperature component (Fig.\,\ref{fig_3vpshockFelines}) fail to reproduce
the data. There is need for lower ionisation stages to explain the excess
observed at $\sim$6.55 keV. This points to the presence of shocked material with
shorter ionisation ages $\tau$. In the framework of the hydrodynamics-based
model from \citetalias{2012ApJ...752..103D}, we find that the main contribution
to the Fe K emission therefore comes from the out-of-plane material
(``\ion{H}{ii} region''), which has temperature and ionisation age producing
emission in the 6.55--6.61 keV range. Another possibility for the low-energy
emission ($\sim6.4$ keV) is fluorescence from near-neutral Fe, including Fe in
the unshocked ejecta. Material in the dense ER clumps, on the other hand, has a
temperature too low to significantly contribute to the line.

Following the evolution of the Fe K line fluxes and centroid energies is crucial
to constrain their origin. Next-generation instrumentation, such as the X-ray
calorimeter aboard Astro-H will be able to resolve lines from different Fe ions,
thus providing even deeper physical insights.

The calibration issues encountered by the \chandra\ team show how important it
is to use \emph{both} observatories to monitor such an important source. To
follow the evolution of the light curve and of the iron lines, subsequent
observations of \snr\ with \xmm\ are highly desired.

\begin{acknowledgements}
The \xmm\ project is supported by the Bundesministerium f\"ur Wirtschaft und
Technologie\,/\,Deutsches Zentrum f\"ur Luft- und Raumfahrt (BMWi/DLR, FKZ 50 OX
0001) and the Max-Planck Society.
P.\,M. and R.\,S. acknowledge support from the BMWi/DLR grants FKZ 50 OR 1201
and FKZ 50 OR 0907, respectively.
\end{acknowledgements}


\end{document}